\documentclass[preprint2]{aastex}

\shorttitle{Molecular hydrogen deficiency in HI poor galaxies}

\begin{document}


\title{Downsizing among disk galaxies and the role of the environment}


\author{Giuseppe Gavazzi \altaffilmark{1}}
\email{giuseppe.gavazzi@mib.infn.it}


\altaffiltext{1}{Departimento di Fisica, Universita' degli studi di Milano--Bicocca, Milano, Italy.}


\begin{abstract}

The study of PopI and PopII indicators in galaxies has a profound impact on our understanding of galaxy evolution.
Their present ($z=0$) ratio suggests that the star formation history of galaxies was primarily dictated by their global mass.
Since 1989 Luis Carrasco and I spent most of our sleepless nights gathering $H_\alpha$ and near infrared 
surface photometry of galaxies in the local Universe and focused most of our scientific career on these two indicators 
trying to convince the community that the mass was the key parameter to their evolution.
We were unsuccessful, until in 2004 the Sloan team rediscovered this phenomenon and named it "downsizing".\end{abstract}

\keywords{Galaxies: evolution -- Galaxies: fundamental parameters}

\begin{figure}[!t]
  \includegraphics[width=1\columnwidth]{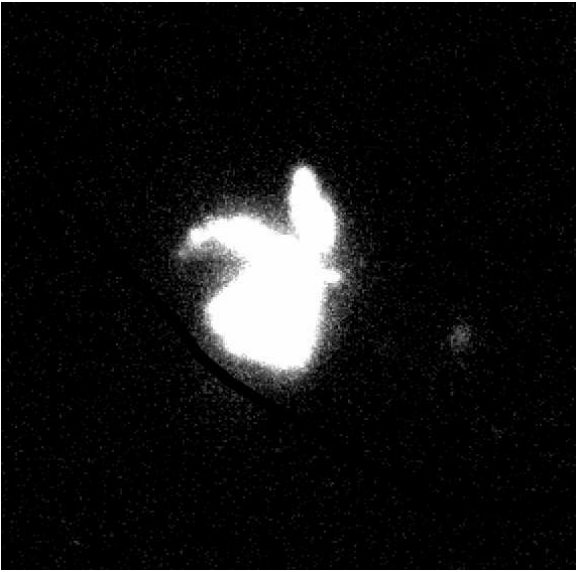}
  \caption{An $H\alpha$ (NET) frame of the Playboy galaxy CGCG 224-038, taken with the 2.1m at SPM.}
  \label{fig:simple1}
\end{figure}

The collaboration with Luis Carrasco dates back to 1989, when we joined our efforts to 
rejuvenate the 2.1m telescope at San Pedro Martir (SPM) by dressing it up with an early
CCD by Photometrics, equipped with narrow-band filters for redshifted $H\alpha$ 
and with the Boller \& Chievens spectrograph, that is still among the instrumentation offered at SPM.\\
Since then our fruitful collaboration brought us to work together on many projects that
eventually gave birth to 13 publications.\\
The first (yet most important) discovery we made quite serendipitously at SPM was that the galaxy CGCG 224-038, which was known
as a starburst, is probably so because it is under the action of "harassment" (see fig. \ref{fig:simple1}).
It was therefore quite natural to baptize it as the "Playboy galaxy", for the glory of Ben Moore (et al. 1999).\\
Later on we started some multifrequency surveying of significant samples of galaxies in the
local Universe, following one of the preferred and most productive activities of Luis. 
We concentrated at obtaining surface photometry in $H\alpha$ and NIR:
the former was aimed at quantifying the massive, ongoing formation of young stars; the latter
the oldest stellar population. The survey is still in progress.\\  
Since the pioneering work of Recillas et al. (1990, 1991) and of Gavazzi et al. (1996) it resulted,
among other things, 
that the NIR band H was providing us with a very robust tool to infer the dynamical mass of galaxies,
up to their optical radii. In other words $M/L$ of spirals was found roughly constant in the infrared,
as opposite to bluer bands.\\
Following the seminal work of Gavazzi et al. (1996), of Boselli et al. (2001) and Scodeggio et al. (2002) 
and exploiting the data-set
that we collected in the past years of collaboration with Luis Carrasco,
in the present paper I wish to put together all possible evidences of
"downsizing" among $disk~galaxies$ \footnote{Let me remind that the word "downsizing" was first introduced
by Cowie et al. (1996)}. In other words I will present and discuss
six independent lines of evidence, e.g. 6 scaling laws, all suggesting that the galaxy mass is the principal
parameter governing the evolution of these systems.
These ideas are receiving some late recognition, as the November 2008 issue of Nature reports
two papers by Disney et al. (2008) and by van den Bergh (2008a) that are focused on these issues and 
confirm our earlier findings.
The very existence of such correlations provides a challenge for hierarchical models of galaxy evolution
(Cole at al. 2000) since, quoting van den Bergh (2008a), "... if galaxies were assembled through the chaotic merging of dark matter halos,
one would expect the properties of individual galaxies to be determined by numerous independent factors, such as
star-forming history, merger history, mass, angular momentum, size and environment...".
The fact that most properties of galaxies seem to scale simply by their mass provides a severe challenge.\\
The six parameters that we will show to correlate (or anti correlate) with the system mass are:
1) the HI mass; 2) the current star formation rate per unit mass; 3) the B-H color index; 
4) the NIR effective surface brightness; 5) the light concentration index and 6) the presence of nuclear features 
in the R band.\\
Studying all correlations we will discriminate between normal and "AGN" hosting systems (including LINERS).\\
Furthermore, aiming at clarifying if the scaling laws are sensitive to the galaxy environment,
all correlations will be shown separately for "normal" (unperturbed) galaxies (left panels) and for 
"perturbed" objects.
The latter discrimination is according to the HI deficiency parameter, that is known 
to provide a sensitive probe of the environmental conditions (Boselli \& Gavazzi 2006).

\section{The data}
\label{sec:sample}
\subsection{The sample}

For the purpose of the present investigation, galaxies have been selected from the Goldmine Database 
(Gavazzi et al. 2003, and references therein) according to the following criteria:\\
1) they belong to the Virgo, Cancer, Coma or A1367 clusters or they lie in the bridge between the
Coma Cluster and A1367;\\
2) they have morphological type Sa or later;\\
3) they have a measurement available for the total magnitude in H band (to provide the dynamical mass);\\
4) they have a measurement available for the integrated HI mass.\\
There are 868 galaxies meeting these requirements. When we divide the sample
according to the HI deficiency parameter (Haynes \& Giovanelli 1986), we find 463 galaxies 
with $HI_{Def} \leq  0.3$ (unperturbed sample) and 251 with $ HI_{Def} > 0.4$ (perturbed sample) 
(the interval $0.3 < HI_{Def} \leq 0.4$ contains 154 objects that have been deliberately omitted).
Needless to say the majority of galaxies with normal HI content belong to the bridge
between Coma and A1367 and to the loose Cancer cluster, while that of HI deficient objects 
is made of members of the two richest clusters: Virgo and Coma.
It should be emphasized that all galaxies, except in the Virgo cluster, are optically selected 
according to the criteria of the CGCG catalogue (Zwicky et al, 1961-68). In the nearby
Virgo cluster the optically selected sample comes from the deeper VCC catalogue of Binggeli et al. (1985)
that is complete down to $m_B=18.0$ mag, thus it includes dwarf Irregular and BCD galaxies.

\subsection{The analyzed parameters} 

The present section lists and describes the parameters that are analyzed in this paper.\\
According to Gavazzi et al. (1996) the dynamical mass (within the optical radius) of disk galaxies
is: $log M_{dyn}=log L_H+0.7$. This quantity is plotted in the horizontal axis of fig.2-7.\\
The classification of galaxy nuclei among normal, AGNs or LINERS is taken from Decarli et al. (2007).\\
The $H_\alpha$ E.W. in fig. \ref{fig:widefig3} determines the intensity of the $H_\alpha$ line over the continuum, thus it
provides a measure of the star formation rate per unit (red continuum) light, i.e.
a quantity that is proportional to the birth-rate parameter $b$, i.e. the present to past star formation ratio
(Kennicutt et al. 1994). The $H_\alpha$ data are from Gavazzi et al. (2006)(and references therein).\\
The NIR mean effective surface brightness $\mu_e$ in fig. \ref{fig:widefig5} is the mean surface 
brightness within the effective radius (Gavazzi et al. 2000).\\
The NIR light concentration index $C_{31}$ in fig. \ref{fig:widefig6} is the ratio of two
radii that contain 75\% and 25 \% of the total light respectively (Gavazzi et al. 2000). It is sensitive to
the presence of central light concentrations, i.e. bulges and nuclei.
Pure exponential disks have $C_{31}\sim 3$, while $C_{31}\geq 3$ is found among galaxies
with significant central enhancements.\\
The nuclearity parameter in fig. \ref{fig:widefig7} is derived from the difference between the mean central surface brightness
within the seeing disk and the surface brightness at the radius that contains 25 \% of the total R band light.
Strong nuclearity parameters ($>1$) are associated with relevant nuclear features in the R band.

\section{analysis}

\begin{figure*}[!t]
  \includegraphics[width=8cm,height=8cm]{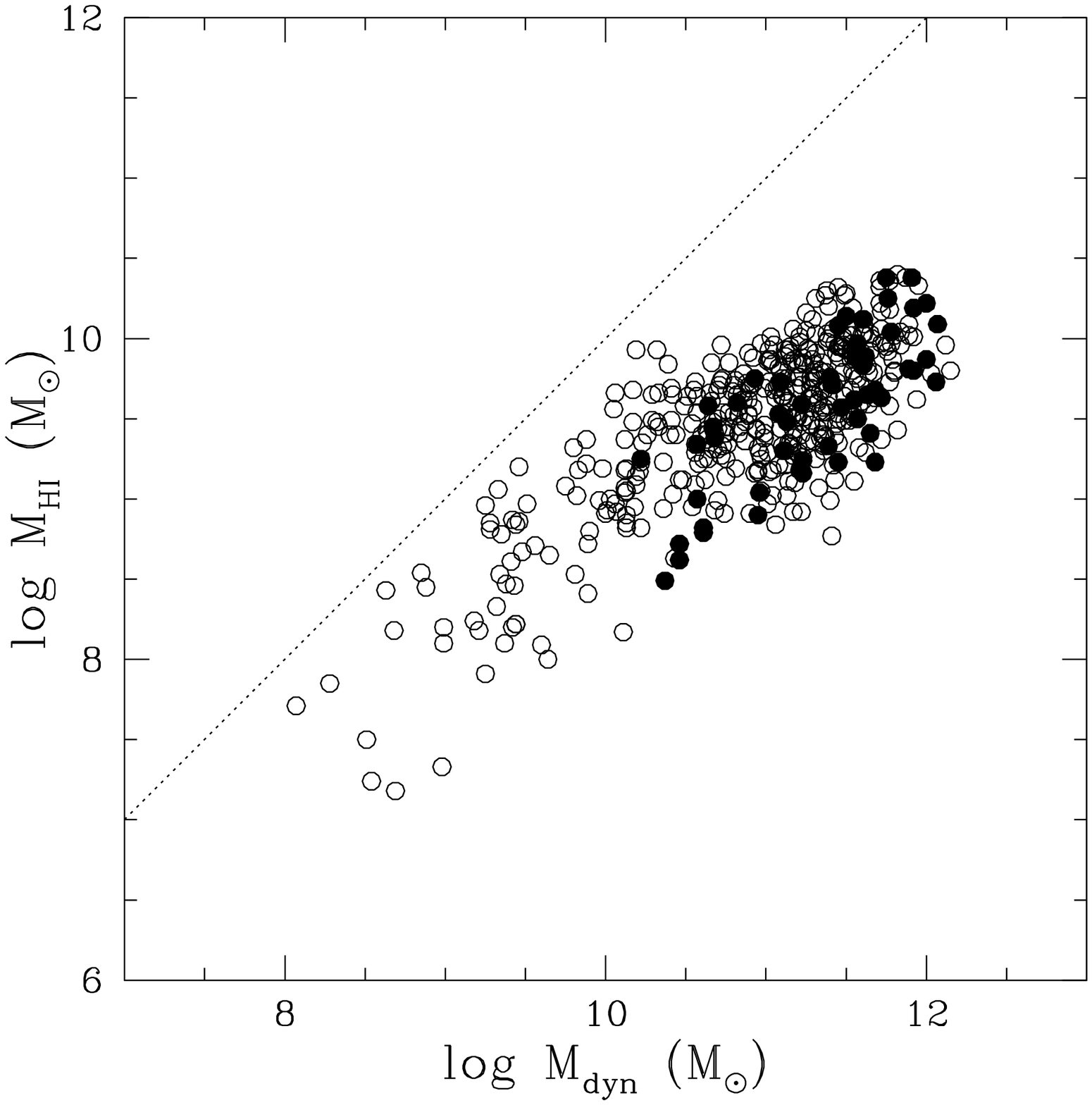}%
  \hspace*{\columnsep}%
  \includegraphics[width=8cm,height=8cm]{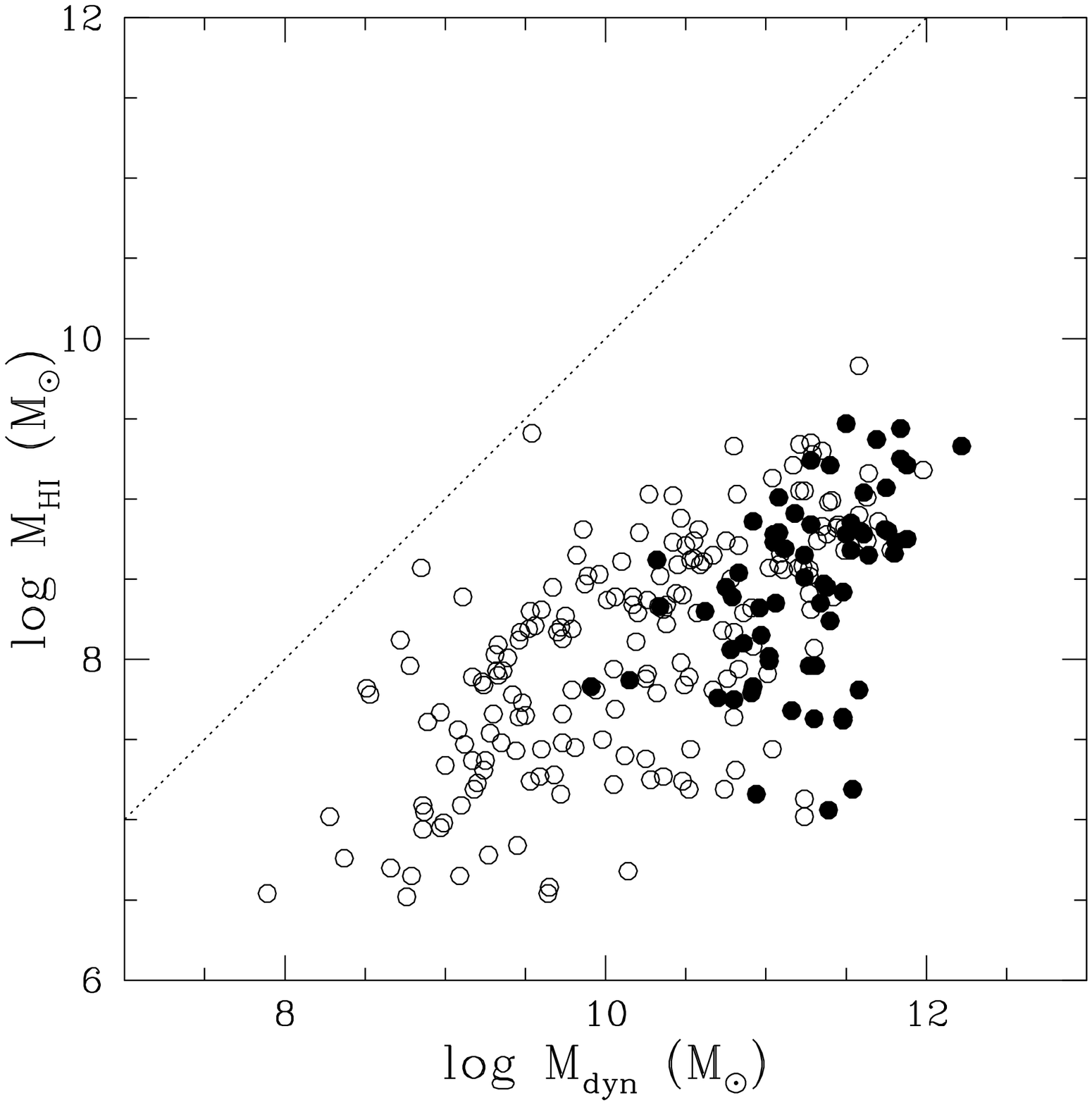}
  \caption{The HI mass vs. dynamical mass relation for normal (Left: $HI_{def} \leq 0.3$) 
 and HI deficient (right: $HI_{def}>0.4$) galaxies. The line represents the one to one relation.
 LINERS and AGNs are filled symbols.}
  \label{fig:widefig2}
\end{figure*}
Fig. \ref{fig:widefig2} (left) shows that the mass of HI gas alone (H2 is neglected here) represents a relevant fraction
(of the order of 15\%) of the total mass for galaxies less massive than $\sim 10^{10} M\odot$ (see also Gavazzi et al. 2008).
Above this value the gaseous fraction becomes progressively smaller with
increasing mass. In this regime a large fraction of galaxies hosts an AGN (Decarli et al. 2007).\\
As expected, among the perturbed (HI deficient) sample (right) there is a significant reduction of the HI mass at any mass,
and the relation becomes a scatter diagram.\\
\begin{figure*}[!t]
  \includegraphics[width=8cm,height=8cm]{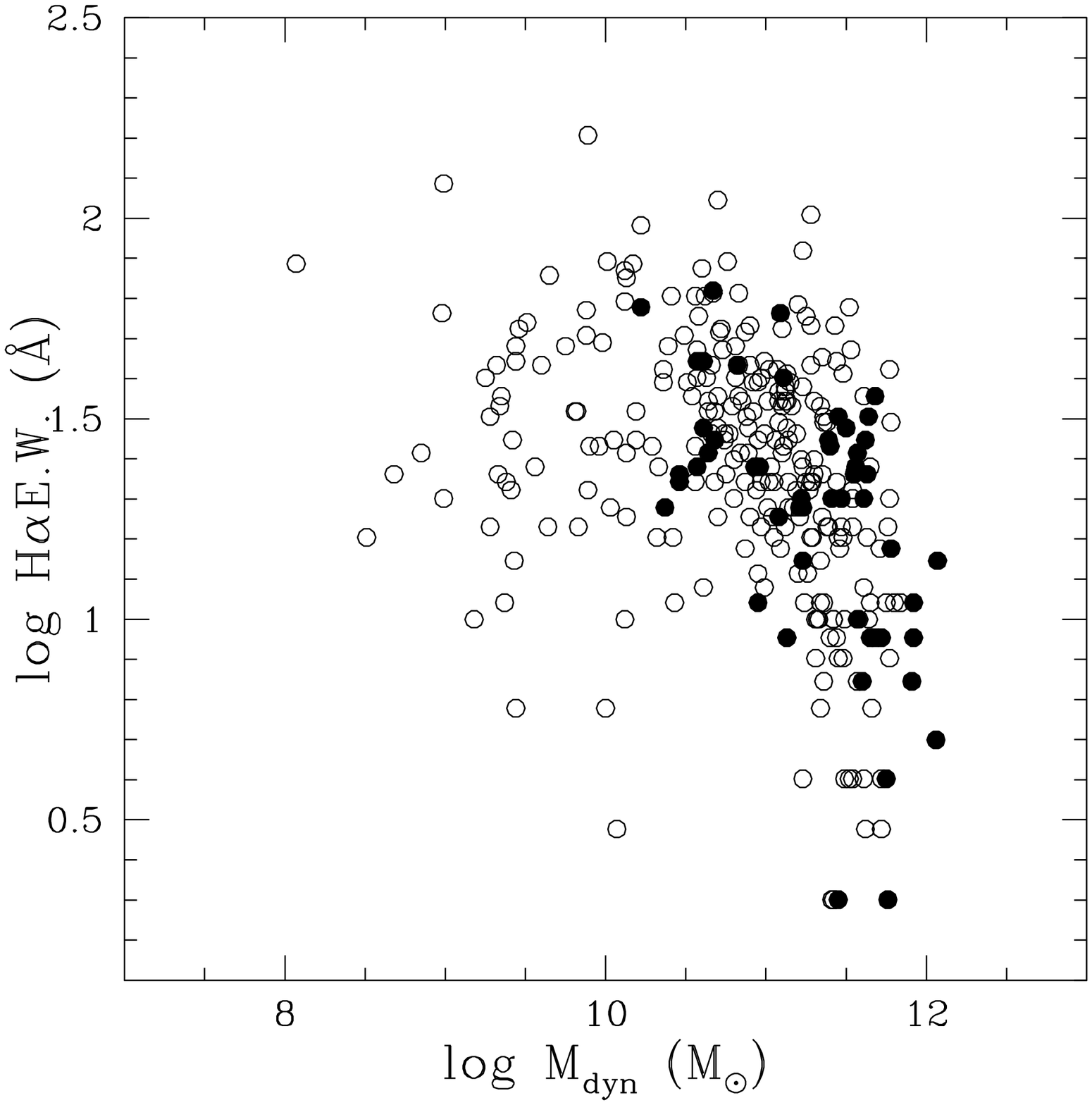}%
  \hspace*{\columnsep}%
  \includegraphics[width=8cm,height=8cm]{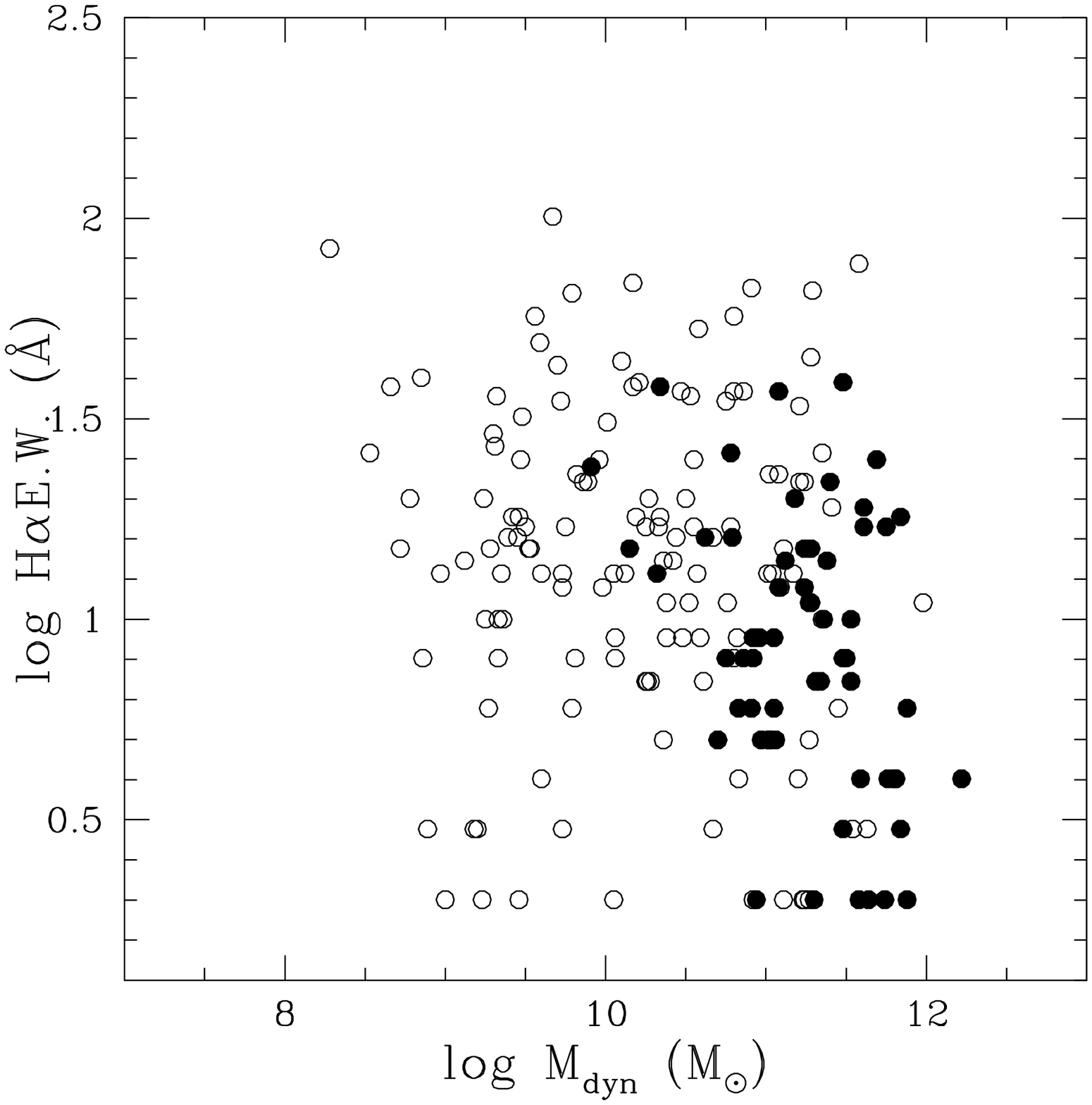}
  \caption{The $H_\alpha$ E.W. vs. dynamical mass relation for normal (Left: $HI_{def} \leq 0.3$) 
and HI deficient (right: $HI_{def}>0.4$) galaxies. LINERS and AGNs are filled symbols.}
  \label{fig:widefig3}
\end{figure*}
Fig. \ref{fig:widefig3} (left) shows that the  massive star formation rate normalized to the continuum luminosity 
shows one order of magnitude scatter for galaxies less massive than $\sim 10^{10} M\odot$.
Above this threshold it decreases significantly with increasing mass (see also Lee at al. 2007). This indicates
that "downsizing" is shaping the Hubble sequence of giant spirals.
Among the perturbed (HI deficient) sample (right) there is a significant increase of the spread and a mean shift
toward lower star formation rate at any mass, indicating that whatever mechanism produces the ablation of primordial 
gas (ram pressure or other interactions), it quenches significantly the star formation of galaxies 
that are members to rich clusters (see also Gavazzi et al. 2008; Fumagalli \& Gavazzi 2008; Boselli \& Gavazzi, 2006).\\
\begin{figure*}[!t]
  \includegraphics[width=8cm,height=8cm]{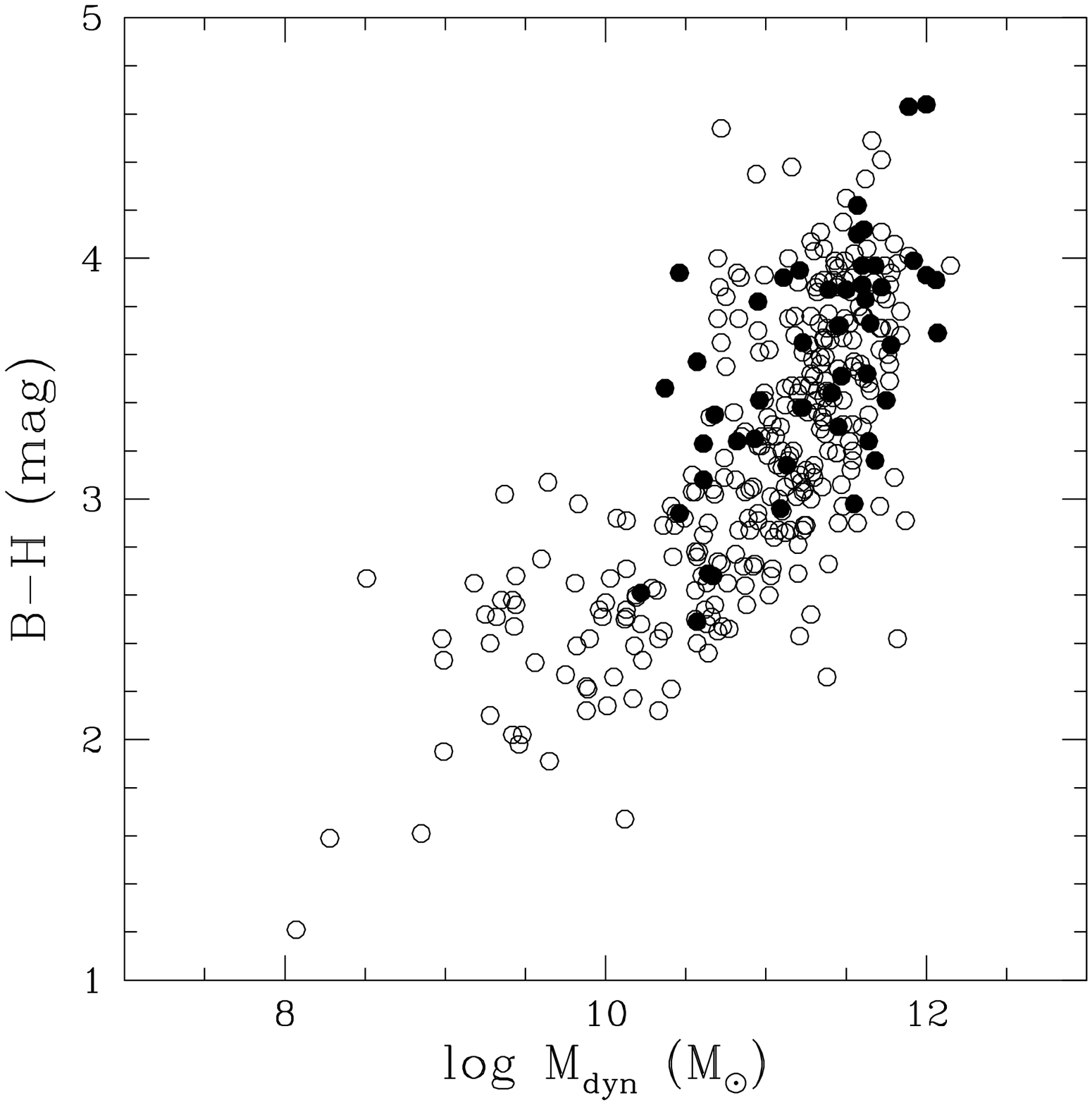}%
  \hspace*{\columnsep}%
  \includegraphics[width=8cm,height=8cm]{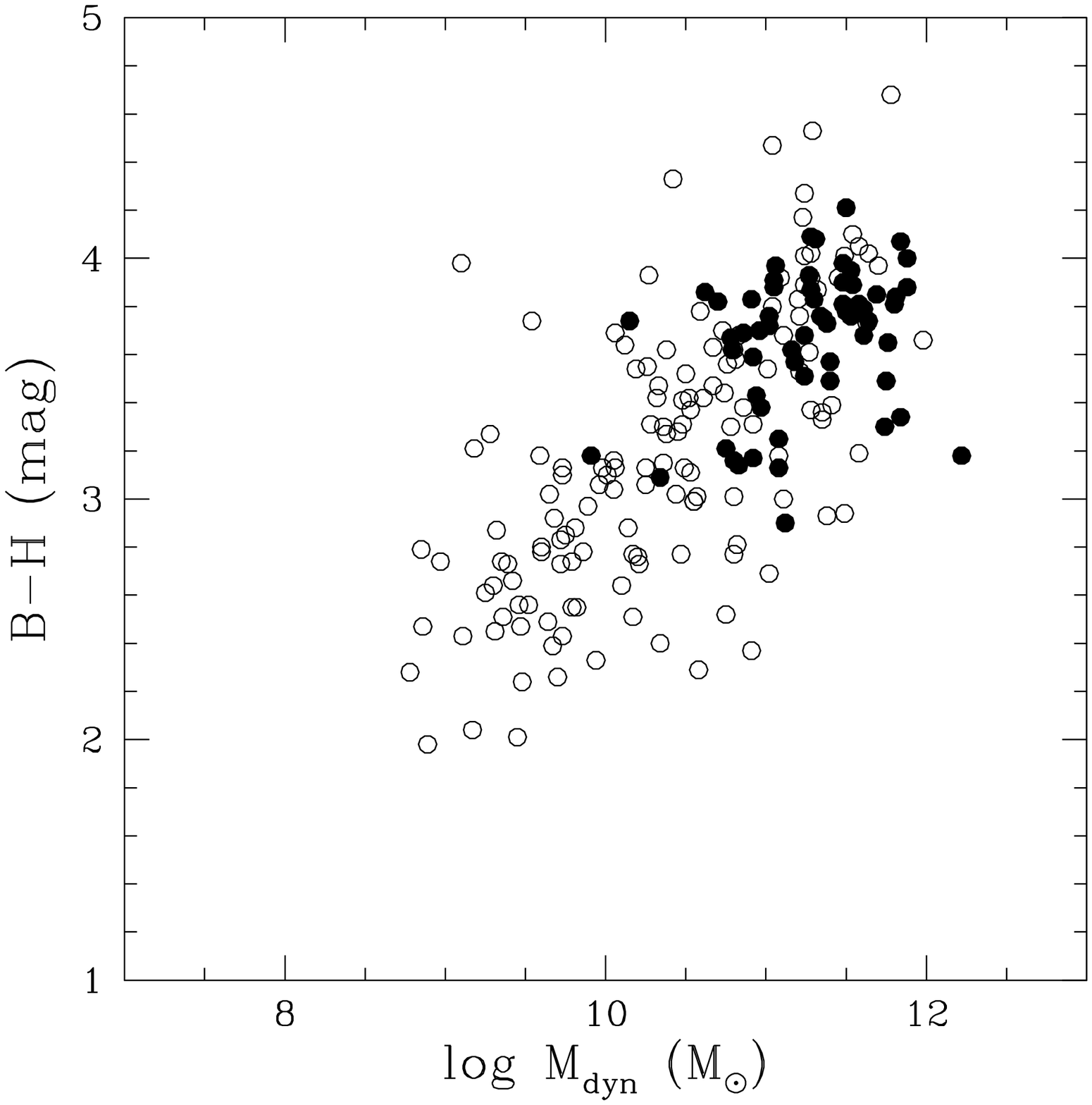}
  \caption{The B-H color index vs. dynamical mass relation for normal (Left: $HI_{def} \leq 0.3$) 
 and HI deficient (right: $HI_{def}>0.4$) galaxies. LINERS and AGNs are filled symbols. }
  \label{fig:widefig4}
\end{figure*}
Fig. \ref{fig:widefig4} (left) shows the (B-H) color-luminosity relation for spiral (and irregular) galaxies with
normal HI content. As opposed to a similar color trend for E+dE galaxies, with a much shallower
slope, that is entirely induced by the metallicity-luminosity and color-metallicity relations,
the color-luminosity effect for spirals spans 4 mag in B-H, 
reflecting a genuine change in population from the dwarfs to the most massive galaxies. 
This effect is consistent with the idea that "downsizing" has been acting for several giga years.
Notice that the color-luminosity relation cannot be attributed to AGN feedback, as proposed by Schawinski et al. (2006)
for early-type galaxies, since
it holds in the whole luminosity interval, including the dwarfs, where AGNs are absent. The same relation for HI deficient galaxies (right) 
shows a larger spread, but is not qualitatively different.\\ 
\begin{figure*}[!t]
  \includegraphics[width=8cm,height=8cm]{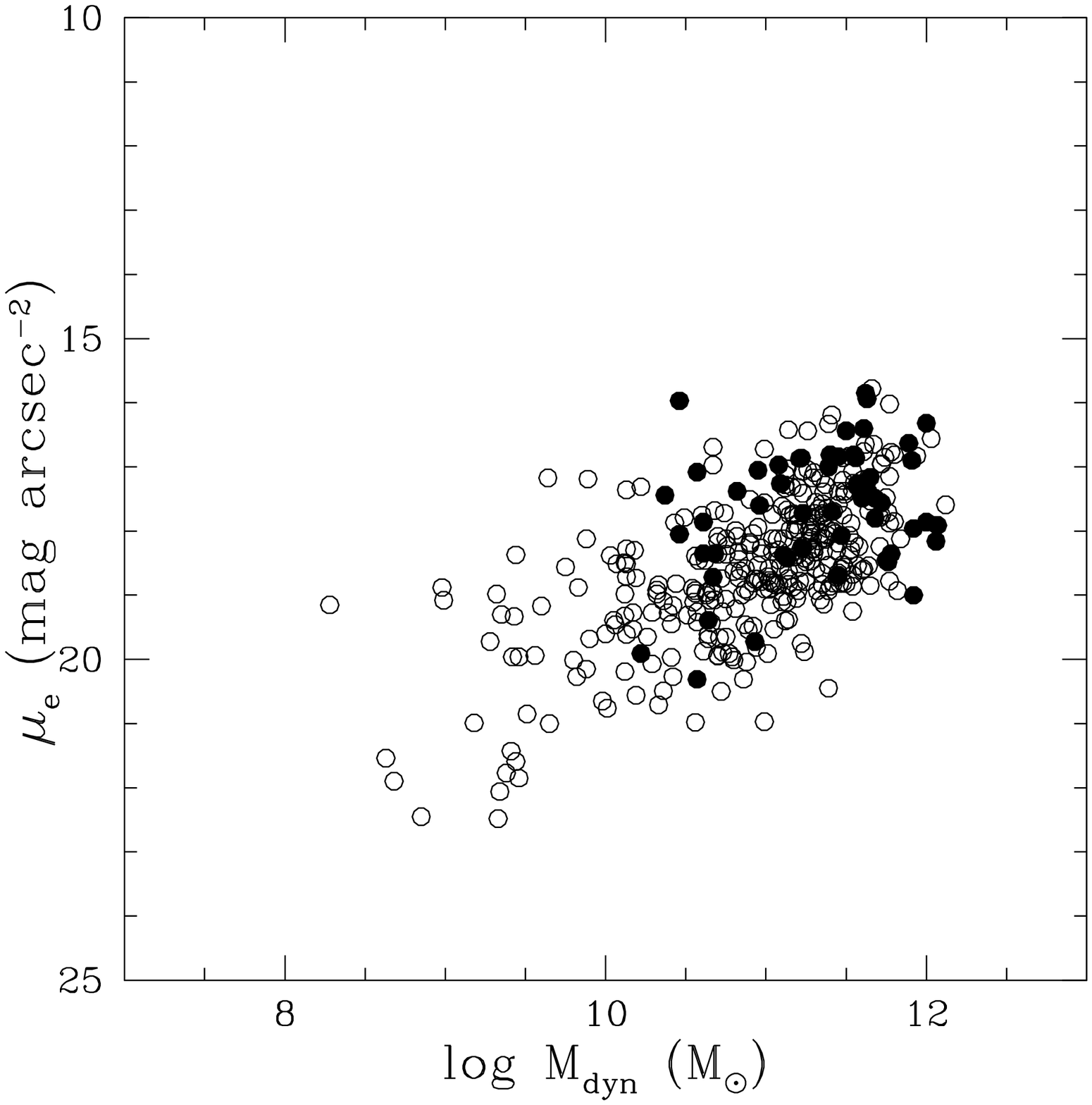}%
  \hspace*{\columnsep}%
  \includegraphics[width=8cm,height=8cm]{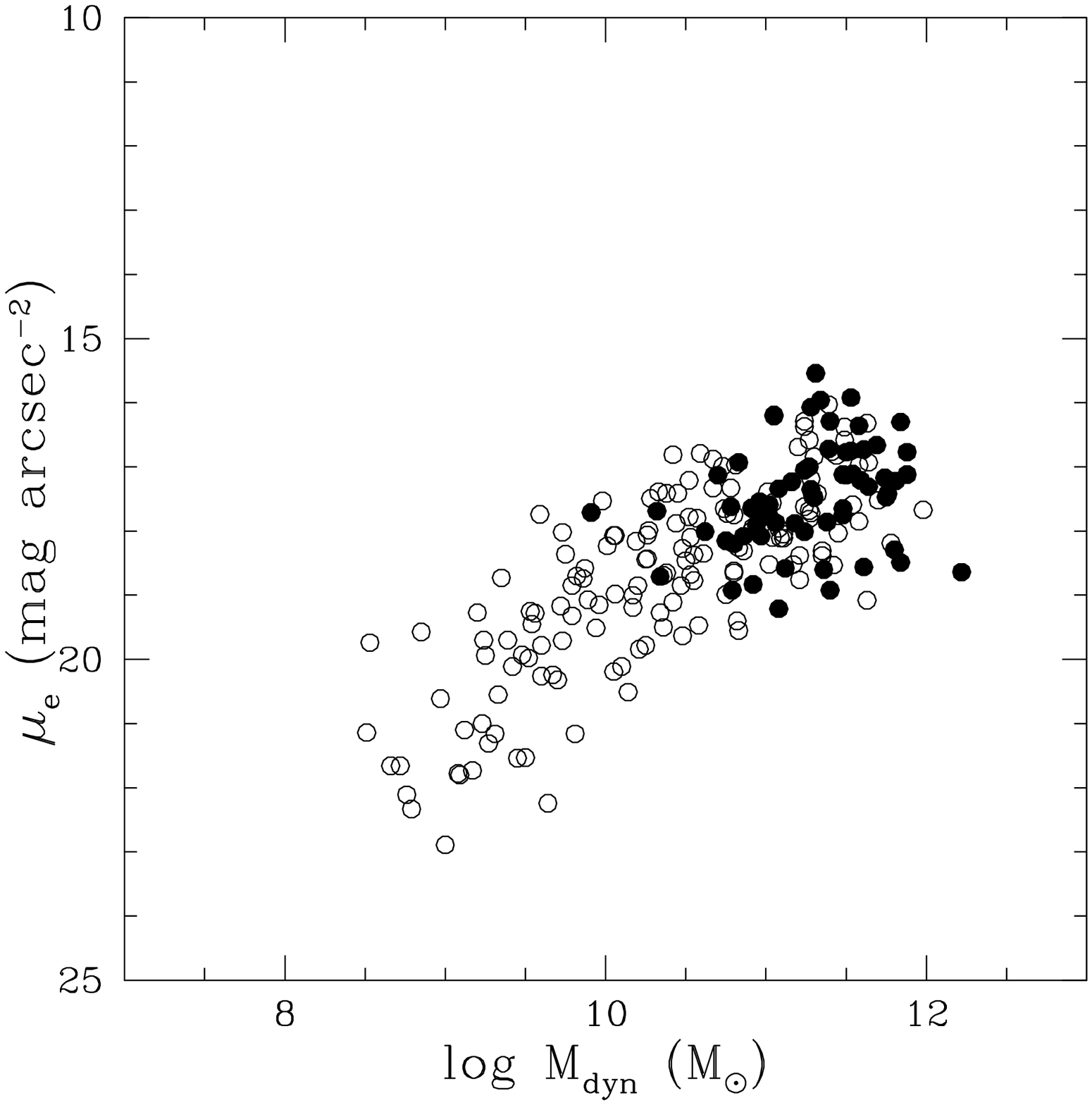}
  \caption{The effective surface brightness vs. dynamical mass relation for normal (Left: $HI_{def} \leq 0.3$) 
 and HI deficient (right: $HI_{def}>0.4$) galaxies. LINERS and AGNs are filled symbols.}
  \label{fig:widefig5}
\end{figure*}
Fig. \ref{fig:widefig5} (left) shows the H band effective surface brightness (mean within $R_e$) vs mass relation.
Notice that this structural scale relation holds identical for normal and deficient (e.g. isolated and cluster) galaxies, 
irrespective of the presence of AGNs at their interior. Old stars are more tightly "packed" in massive galaxies, 
another effect of "downsizing"?\\
\begin{figure*}[!t]
  \includegraphics[width=8cm,height=8cm]{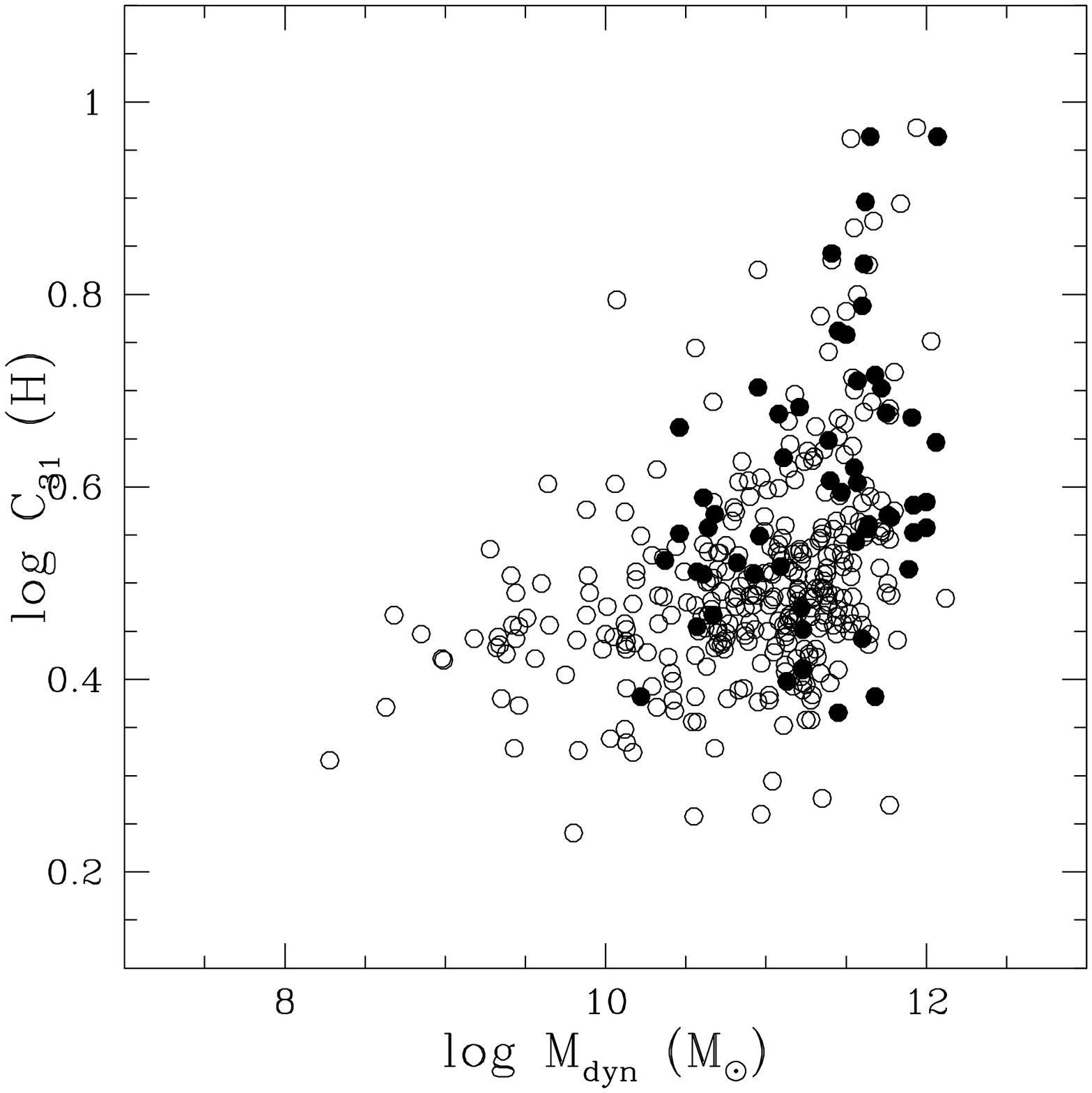}%
  \hspace*{\columnsep}%
  \includegraphics[width=8cm,height=8cm]{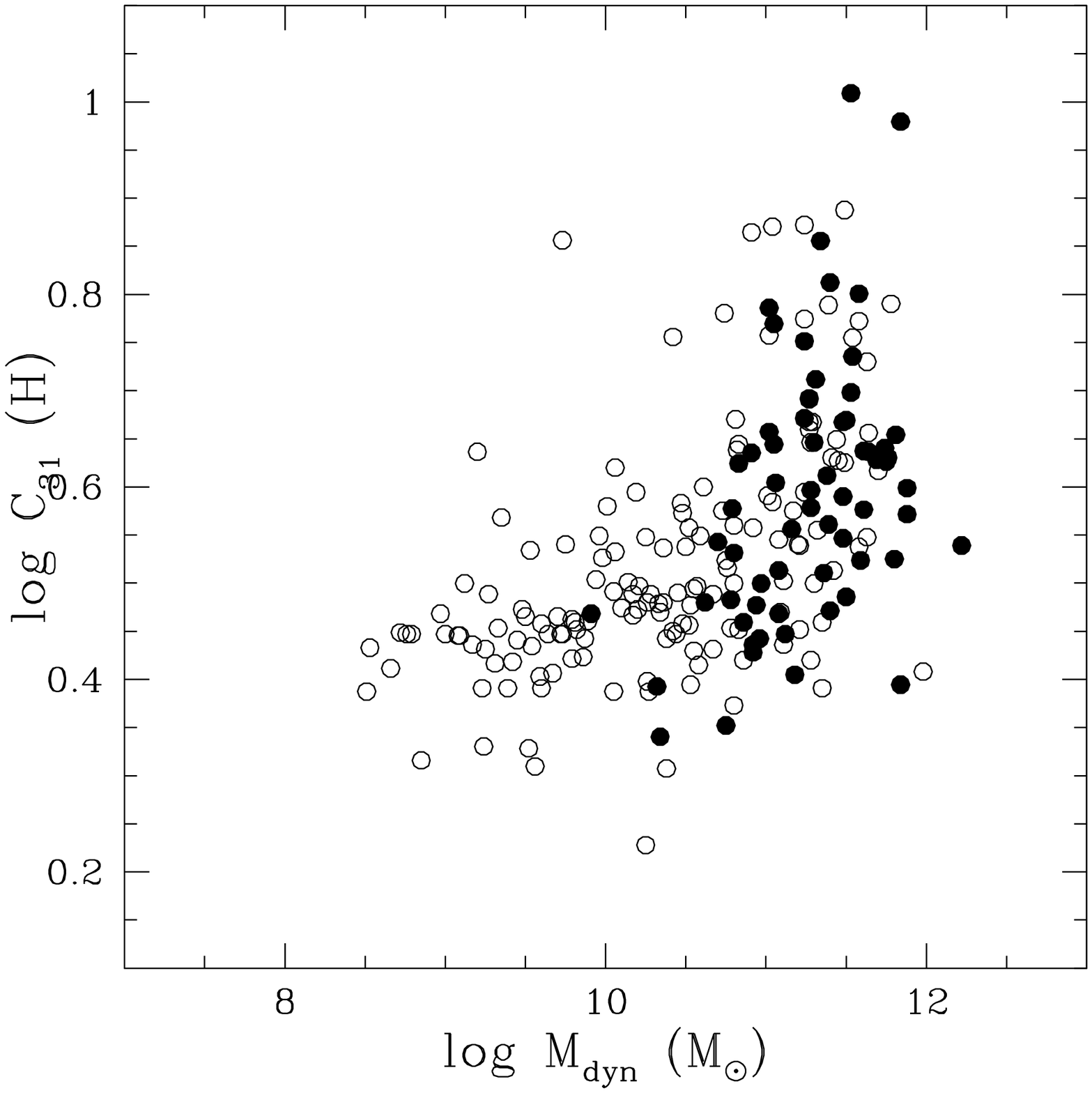}
  \caption{The concentration index vs. dynamical mass relation for normal (Left: $HI_{def} \leq 0.3$) 
 and HI deficient (right: $HI_{def}>0.4$) galaxies. LINERS and AGNs are filled symbols.}
  \label{fig:widefig6}
\end{figure*}
Fig. \ref{fig:widefig6} and Fig. \ref{fig:widefig7} are discussed together, as they are so strongly correlated
one another. The concentration index is in fact sensitive to central light cusps associated with bulge-like 
structures. The nuclearity index is sensitive to smaller nuclear enhancements on the scale of the seeing. 
Both however indicate that the presence of central cusps in disk galaxies is strongly (non-linearly) increasing
with the system mass. Also for this relation the environment seems to play a very marginal role 
(compare left with right panels).\\
\begin{figure*}[!t]
  \includegraphics[width=8cm,height=8cm]{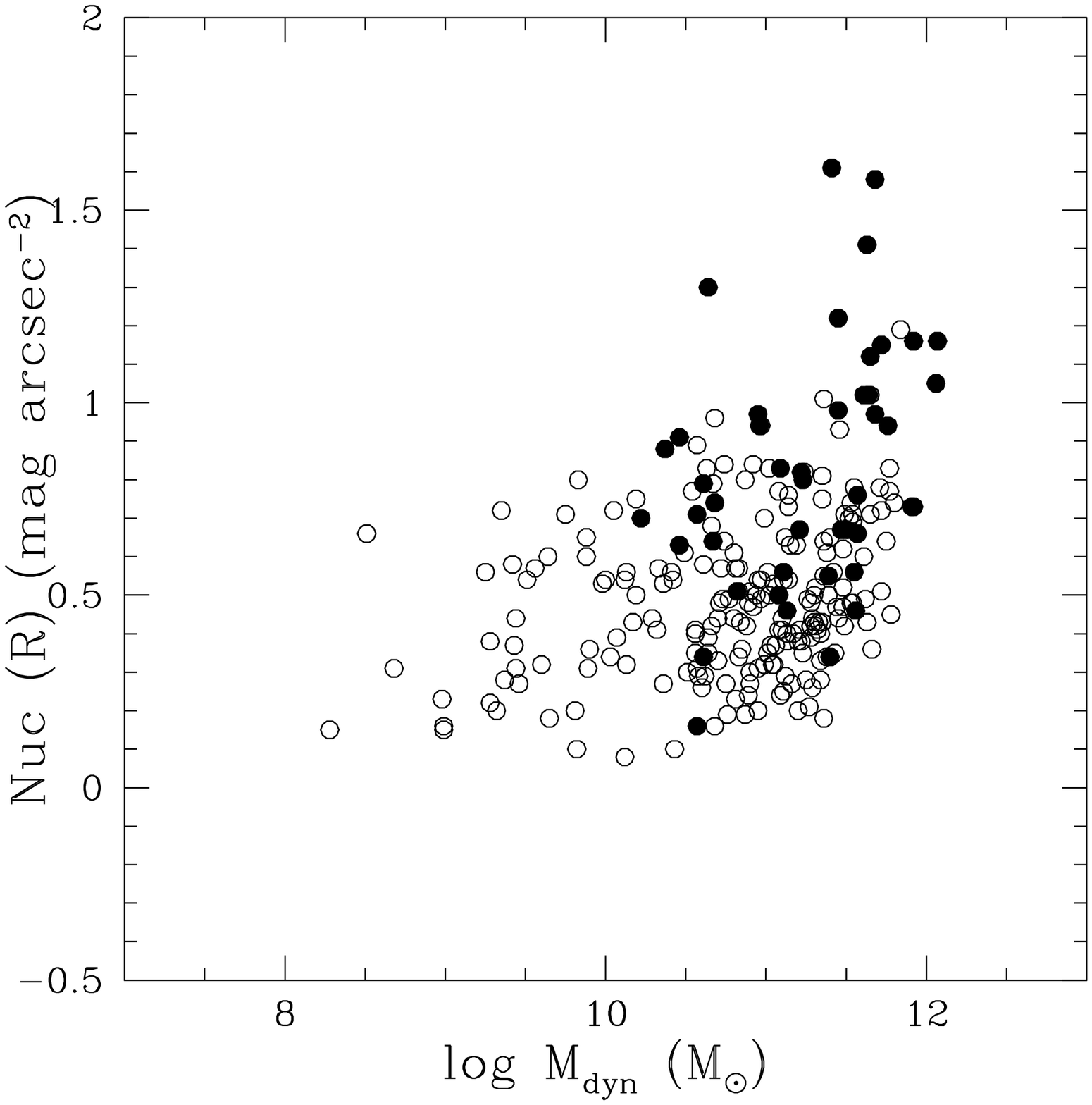}%
  \hspace*{\columnsep}%
  \includegraphics[width=8cm,height=8cm]{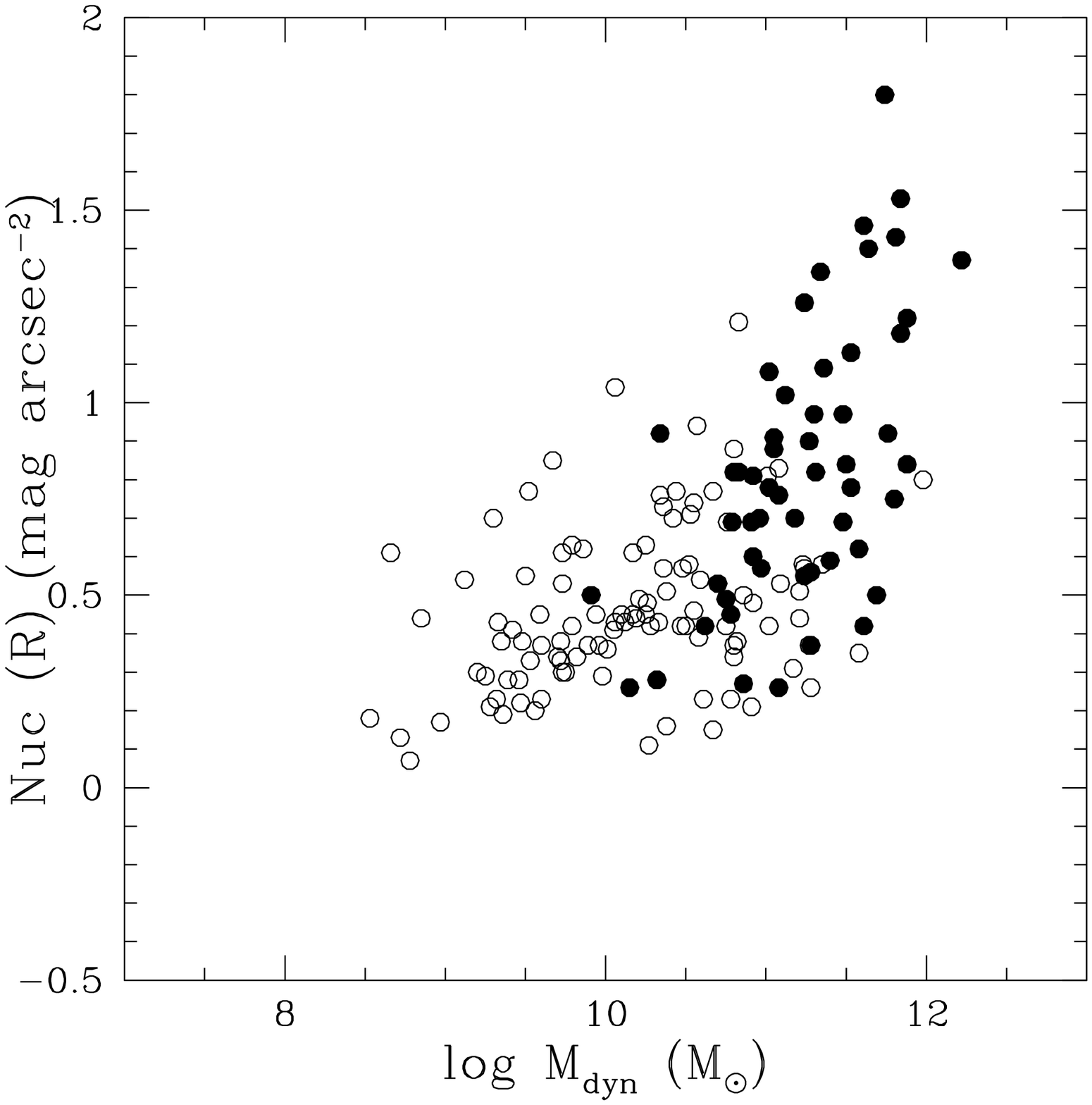}
  \caption{The "nuclearity" in R band vs. dynamical mass relation for normal (Left: $HI_{def} \leq 0.3$) 
and HI deficient (right: $HI_{def}>0.4$) galaxies. LINERS and AGNs are filled symbols.}
  \label{fig:widefig7}
\end{figure*}

\section{discussion and conclusions}

\begin{figure}[!t]
  \includegraphics[width=1\columnwidth]{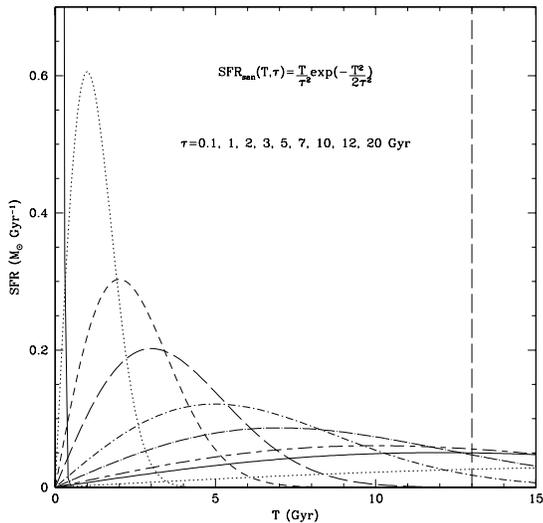}
  \caption{The star formation history as a function of time for various $\tau$ ranging from 0.1 to 20 Gyrs.
  A vertical line is drawn at 13 Gyrs, i.e. at the present epoch.}
  \label{fig:simple8}
\end{figure}

Adopting the principle that scientific models should rely on the "least number of assumptions", 
it is reasonable to conclude that   
the lines of evidence presented in Section 3 find a natural interpretation 
under the simple hypothesis that the star formation history
of disk galaxies is regulated primarily by the system mass, e.g. if the efficiency of protogalaxy
collapse increases with the mass, according to the seminal ideas of Sandage (1986).
Gavazzi et al. (2002) have worked out these ideas by performing SED fitting of disk 
galaxies in the Virgo cluster. They concluded that, adopting families of star formation histories 
that are function of a single parameter ($\tau$), such that their time evolution is as shown in Fig. \ref{fig:simple8},
$\tau$ derived from SED fitting is found inversely proportional to the system mass.
In other words, without invoking  second order feedback mechanism, 
from supernovae and central AGNs, "downsizing" follows naturally assuming that
large mass galaxies have had an earlier, 
shorter and more intense star formation activity than galaxies of progressively lower mass.
This naturally explains why massive systems have reached at the present epoch:\\ 
1) a more complete exhaustion of their HI reservoir (cf. fig. 2);\\
2) higher suppression of the present star formation rate (cf. fig. 3); \\
3) have on average an older stellar population (cf. fig. 4); \\
4) an higher column density of old stars (cf. fig. 5); \\
5) an higher concentration of old stars in the central region (cf. fig. 6-7); \\
The role of the environment, by increasing and speeding up the gas consumption, is obviously producing
a decrease of the present gas content (fig. 2) and consequently the quenching of the current star formation rate
at $z=0$ (fig. 3), otherwise it influences very little other observed structural parameters. \\
In this framework it follows quite naturally that disk galaxies form an almost one-dimensional family in parameter
space (van den Bergh, 2008a,b).   

\end{document}